\documentclass[a4paper,11pt]{article}
\usepackage{jinstpub} % for details on the use of the package, please see the JINST-author-manual
\proceeding{7$^{\text{th}}$ International Conference on Micro-Pattern Gaseous Detectors\\
11–16 December 2022\\
Rehovot, Israel}

\title{\boldmath The novel XYU-GEM to resolve ambiguities}

\author[a,b,1]{K.J. Flöthner \note{Corresponding author.}}
\author[a]{F. Brunbauer}
\author[a]{S. Ferry}
\author[k]{F. Garcia}
\author[a,c,d]{D. Janssens}
\author[b]{B. Ketzer}
\author[a,e]{M. Lisowska}
\author[a,f]{H. Muller}
\author[a]{R. de Oliveira}
\author[a]{E. Oliveri}
\author[a,g]{G. Orlandini}
\author[a,h,i]{D. Pfeiffer}
\author[a]{L. Ropelewski}
\author[h]{J. Samarati}
\author[a]{F. Sauli}
\author[a,f]{L. Scharenberg}
\author[a]{M. van Stenis}
\author[a,j]{A. Utrobicic}
\author[a]{R. Veenhof}

\affiliation[a]{European Organization for Nuclear Research (CERN), 1211 Geneva 23, Switzerland}
\affiliation[b]{Helmholtz-Institut für Strahlen- und Kernphysik, University of Bonn, Nußallee 14–16, 53115 Bonn, Germany}
\affiliation[c]{Inter-University Institute for High Energies (IIHE), Belgium}
\affiliation[d]{Vrije Universiteit Brussel, 1050 Brussels, Belgium}
\affiliation[e]{Université Paris-Saclay, F-91191 Gif-sur-Yvette, France}
 \affiliation[f]{Physikalisches Institut, University of Bonn, Nußallee 12, 53115 Bonn, Germany}
\affiliation[g]{Friedrich-Alexander-Universität Erlangen-Nürnberg, Schloßplatz 4, 91054 Erlangen, Germany}
\affiliation[h]{European Spallation Source ERIC (ESS), Box 176, SE-221 00 Lund, Sweden}
\affiliation[i]{University of Milano-Bicocca, Department of Physics, Piazza della Scienza 3, 20126 Milan, Italy}
\affiliation[j]{Ruder Bošković Institute Bijenička c. 54, 10000 Zagreb, Croatia}
\affiliation[k]{Helsinki Institute of Physics, University of Helsinki, FI-00014 Helsinki, Finland}

\emailAdd{karl.jonathan.floethner@cern.ch}

\abstract{
Removing ambiguities within a single stage becomes crucial when one can not use multiple detectors behind each other to resolve them which naturally is the case for neutral radiation.
An example would be RICH detectors.
Commonly pixilated readout is choosen for this purpose. However, this causes a remarkable increase in quantity of channels and does not scale up well.
Therefore, the XYU-GEM was proposed as a three coordinate strip-readout which is combined with a triple GEM detector.
The readout complements a common XY readout with an additional projection which is tilted by 45°.
The overdetermination due to three projections can be used to resovle ambiguities.
Following the detector design will be explained, first measurements discussed to understand the response of the detector and a way to change the charge sharing without changing the manufacturing parameters of the readout.
}


\begin{document}
\maketitle
\flushbottom

\section{Introduction}
\begin{figure}[h!]
\centering
\label{fig:Closeup}% \begin{center}/\end{center} takes some additional vertical space
\includegraphics[width=.6\textwidth]{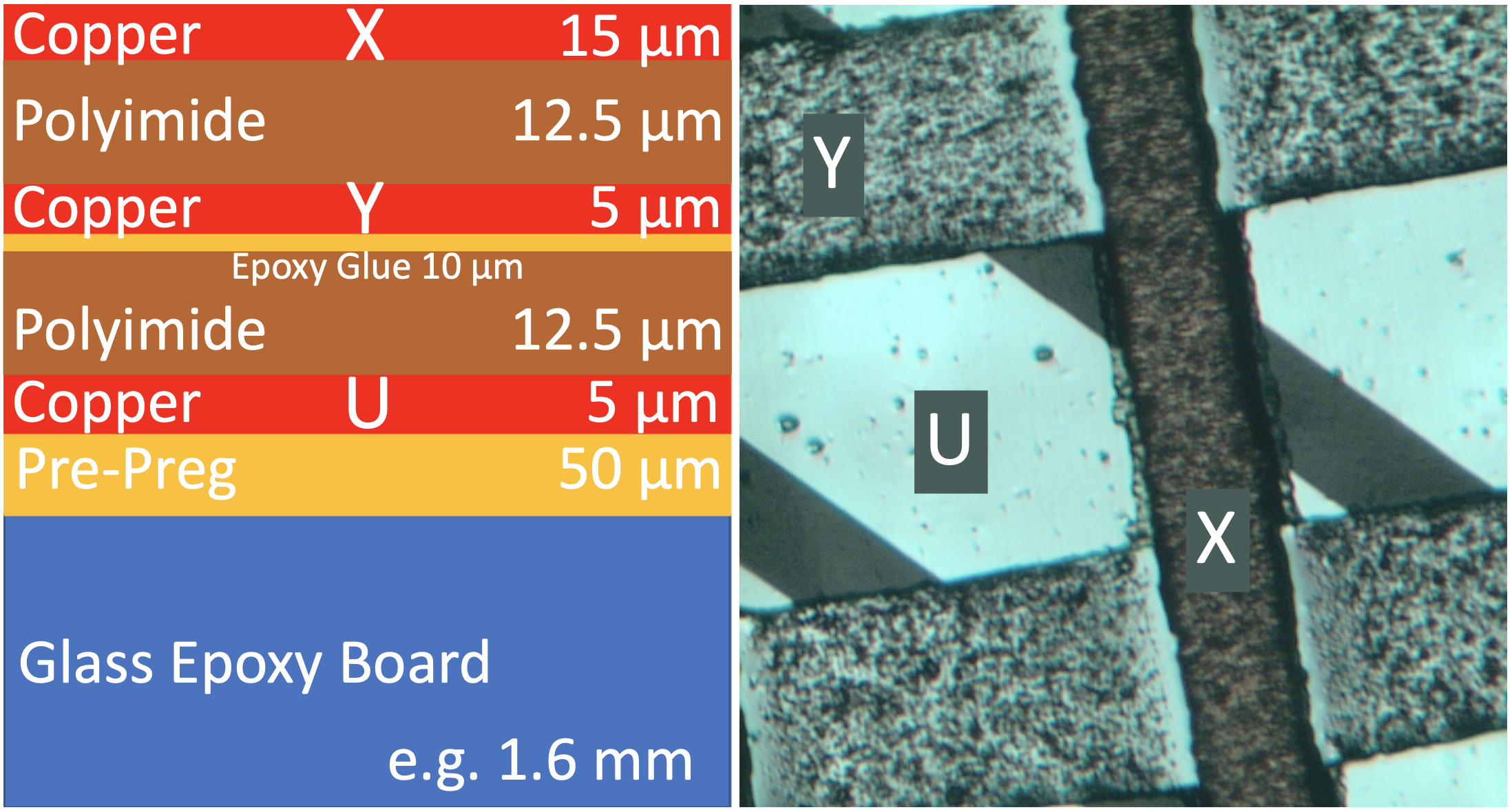}
% "\includegraphics" from the "graphicx" permits to crop (trim+clip)
% and rotate (angle) and image (and much more)

\caption{Left: Schematic cross section. Right: Microscope picture of the three strip layers.} 
\end{figure}

Within limits, ambiguities can be resolved by using electronics which give access to better time resolution or additional information about signal amplitude.
However, solutions can be found as well at detector level by specific designs~\cite{SAULI200518}\cite{Laktineh}.
The proposed XYU readout would be such a solution which does not depend on the used electronics.
A cross-section and microscope picture of the proposed XYU readout can be seen in fig.~\ref{fig:Closeup}.

\begin{figure}[h!]
\centering
\includegraphics[width=.42\textwidth]{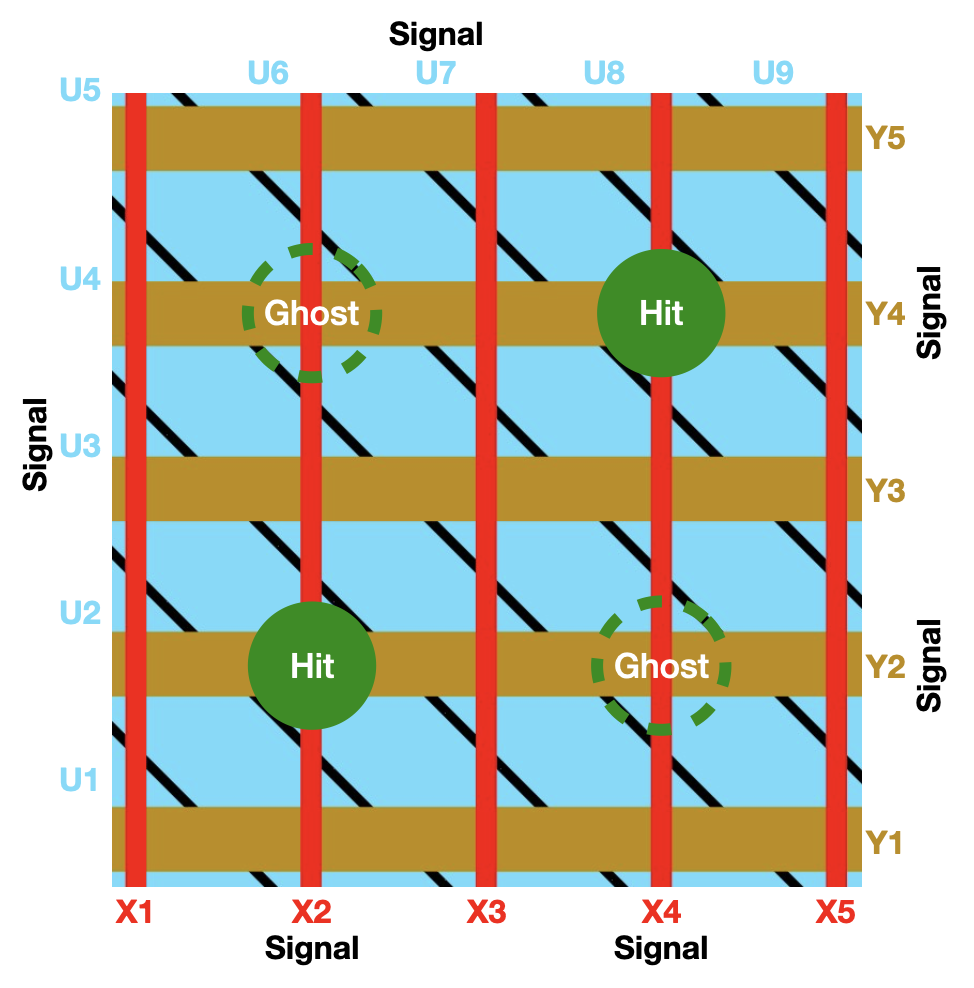}
\qquad
\includegraphics[width=.38\textwidth]{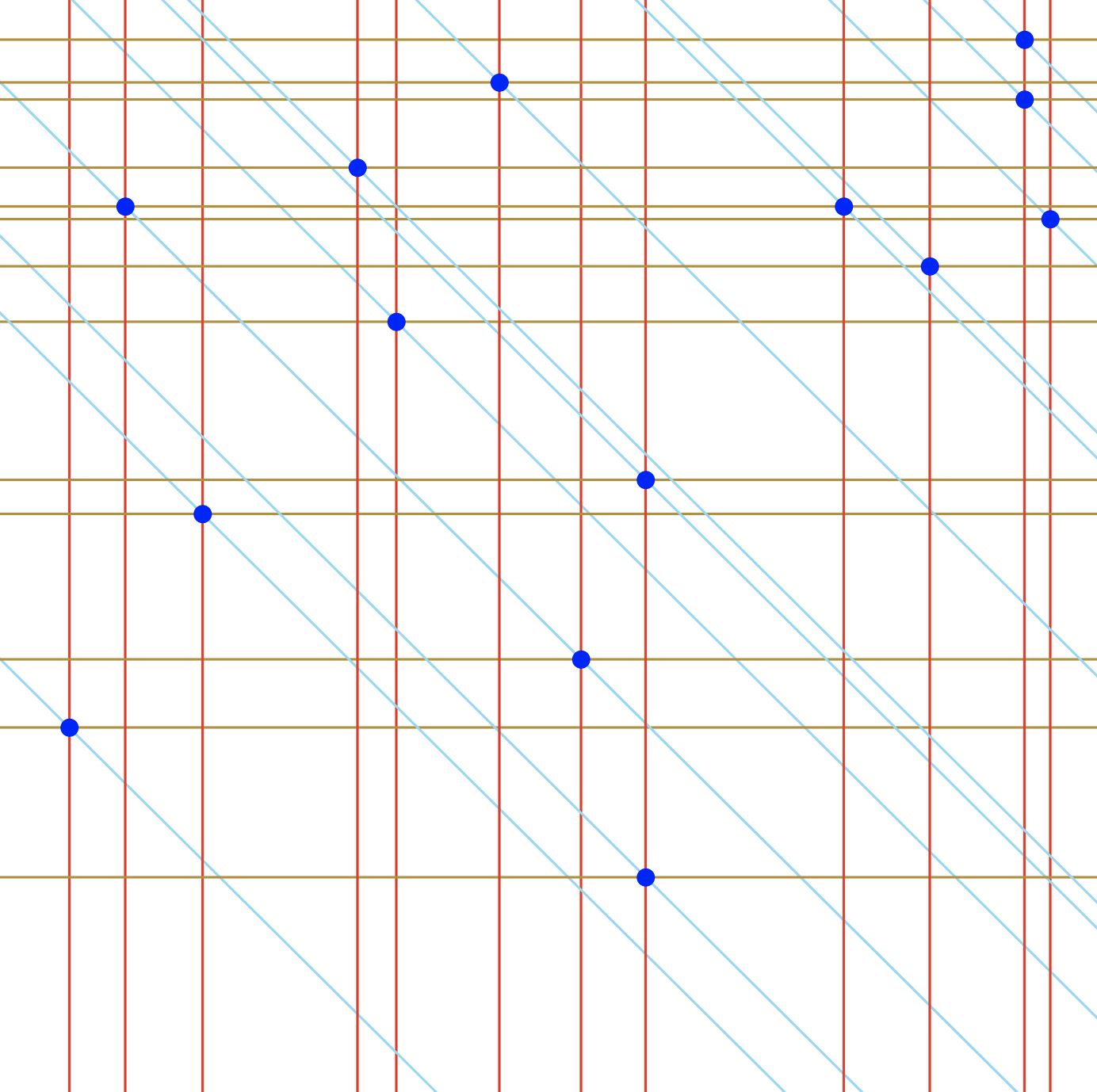}
\caption{Left: Schematic of ambiguities. Right: Illustrated Multi-hit example.
\label{fig:Cap}}

\end{figure}

How the three coordinates can be used to resolve ambiguities is shown in fig.~\ref{fig:Cap}.
As depicted on the left it is not possible to resolve ambiguities with a binary readout with only using two projections. But obvious with the third coordinate. 
On the right resolving most of the ambiguities by simple correlation is already possible.
Still present ghosts might be resolved by more complex combinatory algorithm, e.g. looping over the multiple possibilities and requiring to use up all signals.
A multi-hit event like illustrated in~\ref{fig:Cap} would be impossible to reconstruct with two projections, while one would be able resolve this using the third projection.
This has to be verified with actual multi-hit events, caused by showers, from test beam data.

\section{Detector Design and Setup}
\label{sec:intro}
From simulations a smaller charge collection for U is predicted for the current parameters which are limited by manufacturing constrains~\cite{XYU}. In principle it would be desired to have an equal sharing to avoid the need of operating the detector at higher gains to achieve full efficiency with all projections.
These simulations were based on the Ramo-Shockley theorem~\cite{Shockley}\cite{Ramo} and can provide the total induced charge on the different strip layers.
The picture in fig.~\ref{fig:Closeup} shows all three conductive layers of the readout which are exposed to the gas volume. On top visible the X strips, by 90° tilted the Y strips and by another 45° the U strips.
The manufacturing process was developed by the CERN Micro Pattern Technology (MPT) group~\cite{Rui}.
The process involves the glueing of two copper-cladded polyimide foils, hosting the different strip-layers.
To reveal projections, three dielectric etchings need to be performed.
The result can be seen in fig.~\ref{fig:Closeup} as it was stated above.
The detector itself is based on the standard COMPASS-like triple-GEM detector~\cite{ALTUNBAS2002177}.
The gas mixture Ar/CO$_2$ 70/30 was used and an passive voltage divider (R in M$\Omega$: 1/0.55/1/0.5/1/0.44/1).
For the readout electronics the VMM3a ASIC with the RD51 Scalable Readout System (SRS) was used~\cite{LUPBERGER201891}.

\section{General response of the Detector}
\begin{figure}[h!]
\centering
\label{fig:spectra}
\includegraphics[width=.42\textwidth]{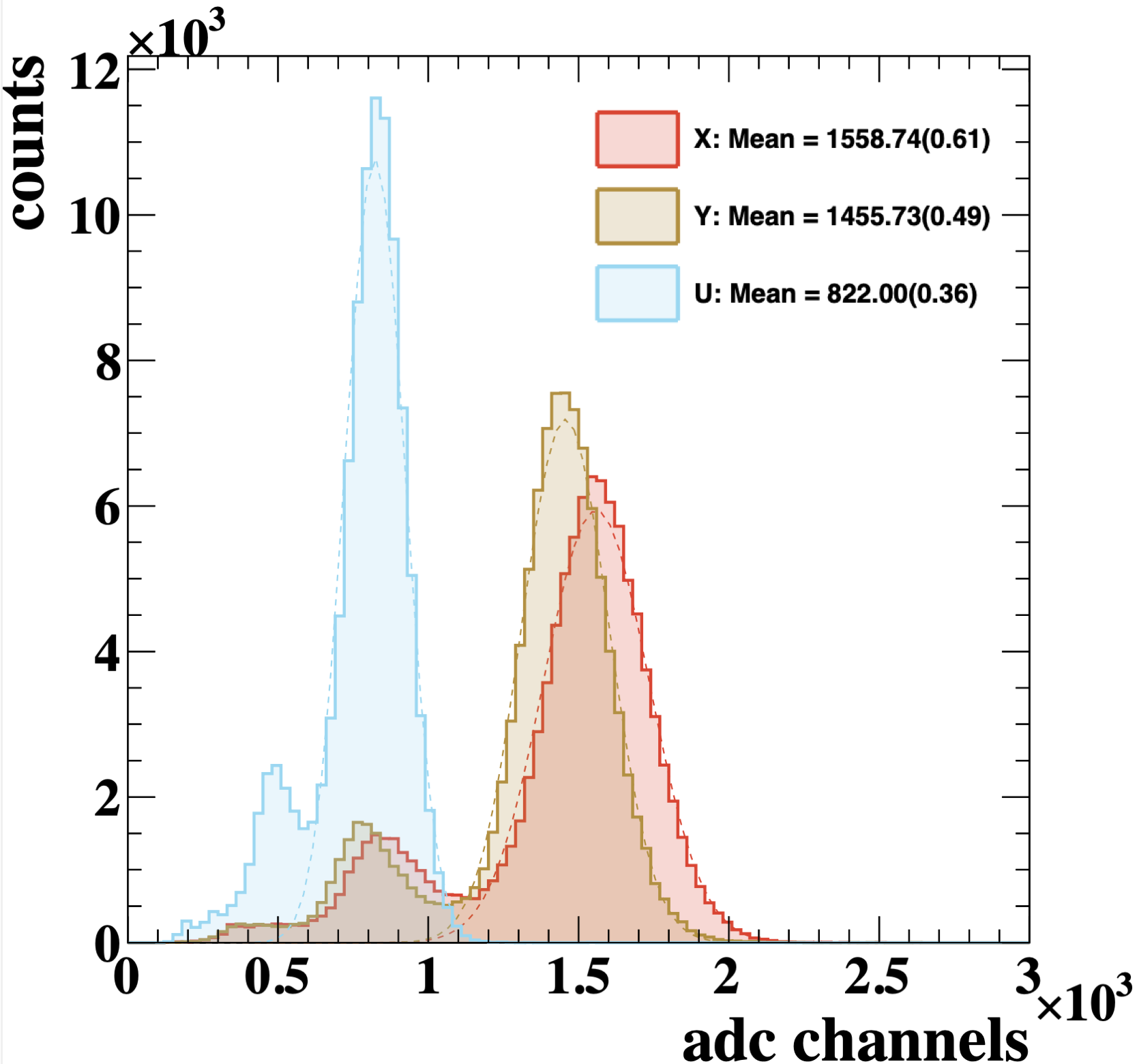}
\qquad
\includegraphics[width=.42\textwidth]{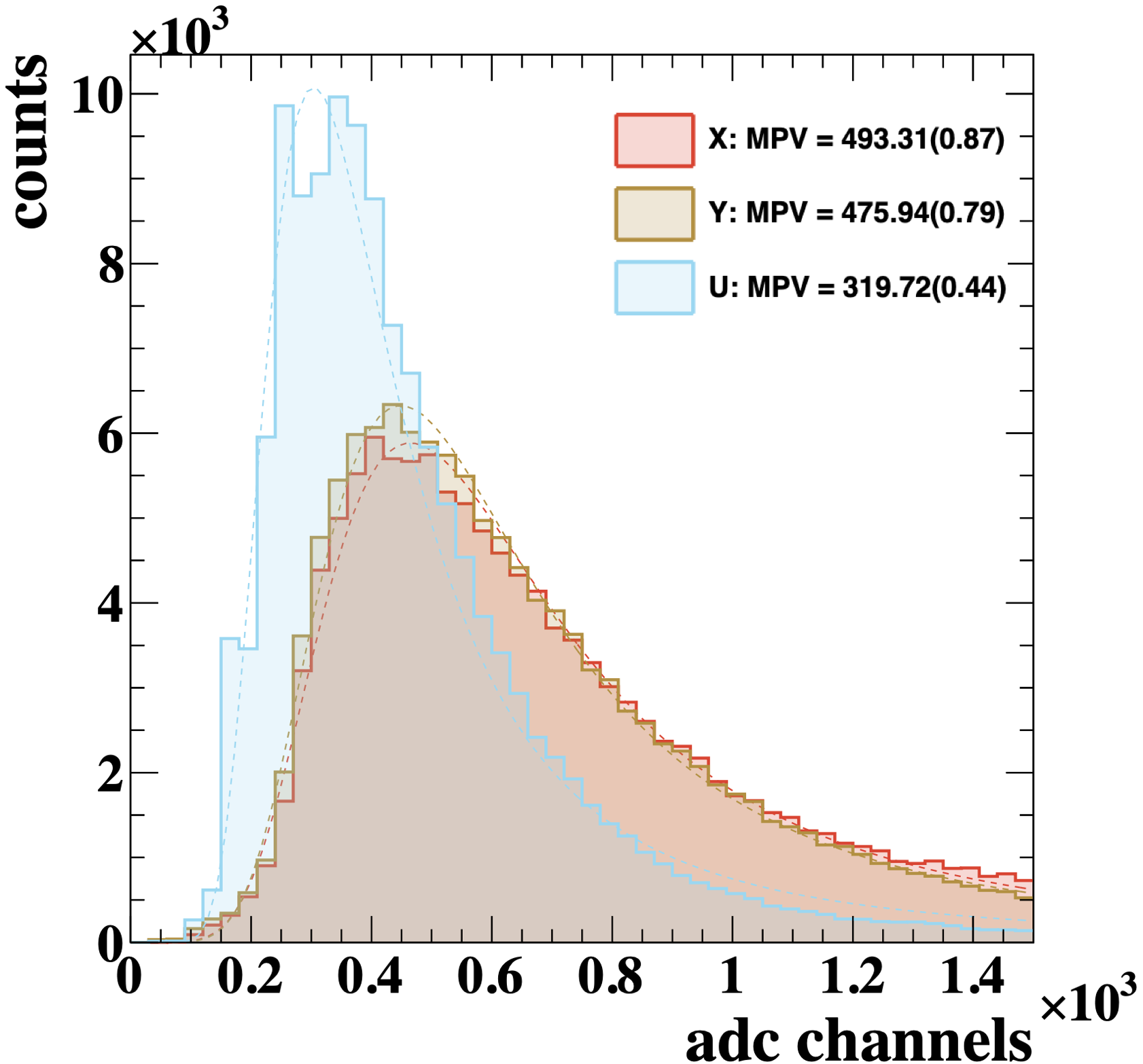}
\caption{Left: ADC spectra for $^{55}$Fe. Right: ADC spectra for 150 GeV/c muons}
\end{figure}

The response of the XYU detector was investigated with $^{55}$Fe in the laboratory and with Minimum Ionizing Particles (MIPs) during the RD51 October test-beam at the SPS H4 beam-line.
Due to the lower energy deposition from MIPs, compared to $^{55}$Fe, the gain for the test-beam  is increased by a factor of three\footnote{The gain during laboratory measurements was \textasciitilde10k.}. 
As shown by the ADC counts, fig.~\ref{fig:spectra}, the distribution looks similar for all three coordinates. Altough one can observe a shift to lower adc channel for the U coordinate, which indicates the lower charge collection. 
Extracting the position of the photopeak and the MPV, the charge sharing can be determined.
This leads to ~41/38/21 (X/Y/U)\footnote{A minor change of sharing with radiation can be observed as well caused by charging up of the insulator~\cite{XYU}.} for the $^{55}$Fe and ~39/36/25 for the beam measurements, which matches with the simulations~\cite{XYU}.

\section{Biasing Strips and Charge Sharing Variation}
\begin{figure}[h!]
\centering % \begin{center}/\end{center} takes some additional vertical space
\includegraphics[width=.48\textwidth]{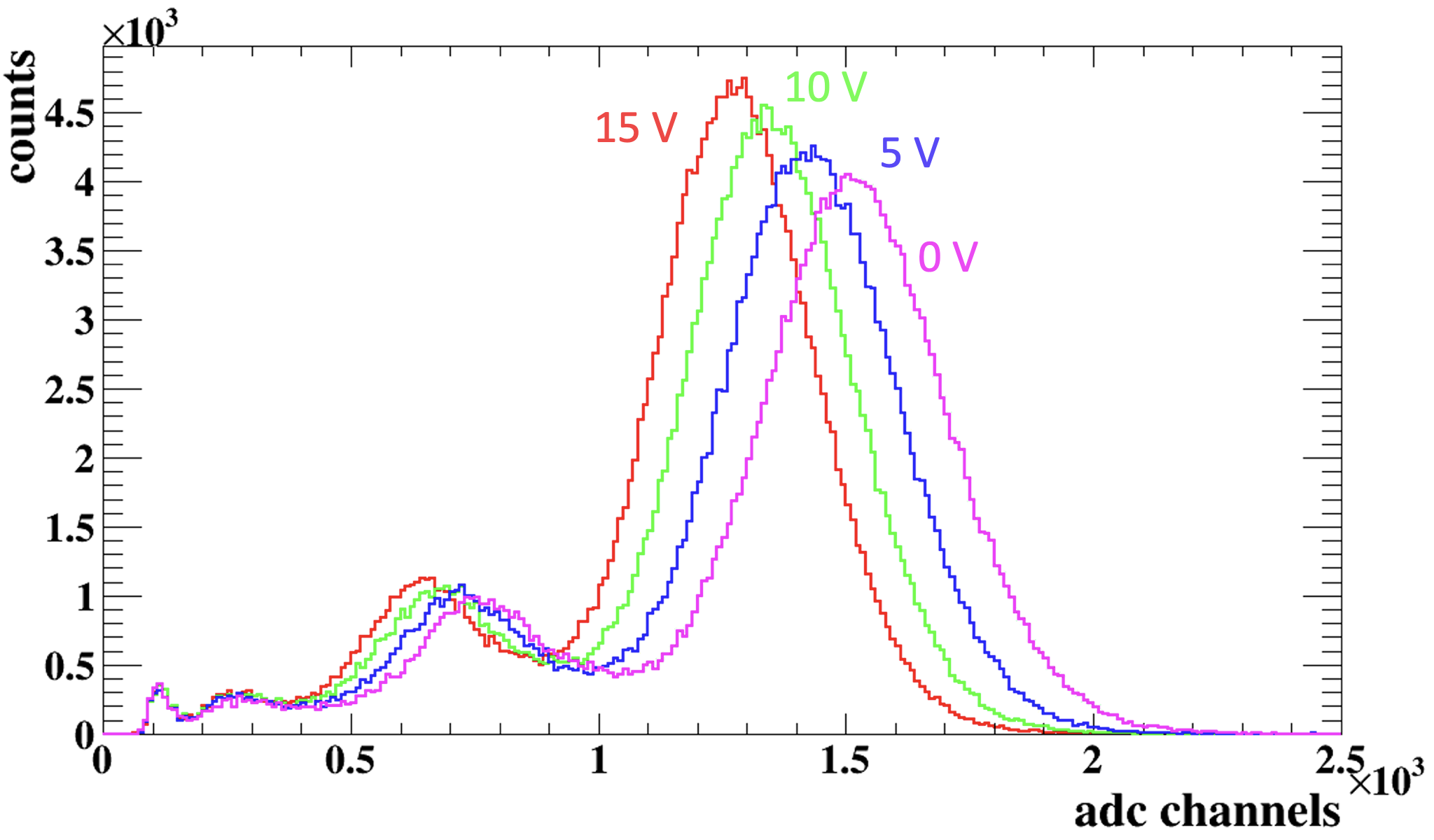}
% "\includegraphics" from the "graphicx" permits to crop (trim+clip)
% and rotate (angle) and image (and much more)
\caption{Spectra from the X strips for different bias voltages on U strips.}
\label{fig:Biasing} 
\end{figure}

Applying a voltage on the strips can be used to compensate the unequal charge collection.
As shown in fig.~\ref{fig:Biasing} one can see a clear shift of the spectrum collected from the X strips while applying different voltages on the U strips.
Therefore one could achieve equal sharing without changing the manufacturing parameters.

\section{Summary and Outlook}
The first prototype of the XYU-GEM could operated and it is possible to collect reasonable data using an $^{55}$Fe source and 150 GeV/c muons. The observed characteristics are as expected.
It is shown, that one can influence the sharing of the signal by applying a bias voltage on the strips. However, further investigations are needed to see the effect simultaneously on all projections.
Using the beam data the detector response with actual multi-hit events, caused by showers, is under investigation.
As well as the possible influence of the third coordinate on the resolution.
Another aspect worth to investigate would be a different pitch of the U projection. This can be done merging channels and therefore changing the signal-to-noise-ratio on individual strips which could be compared to simulations.

%\paragraph{Note added.} This is also a good position for notes added after the paper has been written.
\section*{Acknowledgments}\label{sec: Acknowledgments}
This work has been sponsored by the Wolfgang Gentner Programme of the German Federal Ministry of Education and Research (grant no. 13E18CHA).
The work has been performed in the context of the CERN Strategic Programme on Technologies for Future Experiments. \url{https://ep-rnd.web.cern.ch/}

% Bibliography

%% [A] Recommended: using JHEP.bst file
\bibliographystyle{JHEP}
\bibliography{biblio.bib}

%% or
%% [B] Manual formatting (see below)
%% (i) We suggest to always provide author, title and journal data or doi:
%% in short all the informations that clearly identify a document.
%% (ii) please avoid comments such as "For a review'', "For some examples",
%% "and references therein" or move them in the text. In general, please leave only references in the bibliography and move all
%% accessory text in footnotes.
%% (iii) Also, please have only one work for each \bibitem.

%\begin{thebibliography}{99}

%\bibitem{a}
%Author,
%\emph{Title},
%\emph{J. Abbrev.} {\bf vol} (year) pg.

%\bibitem{b}
%Author,
%\emph{Title},
%arxiv:1234.5678.

%\bibitem{c}
%Author,
%\emph{Title},
%Publisher (year).

%\end{thebibliography}
\end{document}